\documentclass[12pt,a4paper]{article}
\usepackage{amsfonts,amsmath,amssymb}
\vspace{20mm}

\begin{document}

\title{\LARGE \bf Models of relativistic particle with curvature and torsion revisited}

\author{Josu Arroyo, Manuel Barros and Oscar J.Garay}
\date{}
\maketitle

\begin{abstract}
\noindent Models, describing relativistic particles, where Lagrangian densities depend linearly on
both the curvature and the torsion of the trajectories, are revisited in $D=3$ space forms. The
moduli spaces of trajectories are completely and explicitly determined using the Lancret program.
The moduli subspaces of closed solitons in the three sphere are also determined. \\

\vspace{2 mm}

\end{abstract}

\section{Some background}

\noindent The conventional approach to consider Lagrangians that describe relativistic particles is
based on certain extensions of the original space-time by extra variables that provide the required
new degrees of freedom. Recently however, a new approach appeared in the literature (see for
example \cite{Arreaga-Capovilla+, Kuz-Ply, Nersessian++, Nesterenko++1, Nesterenko++2, Ply1, Ply2,
Ply3} and references therein). In this setting, the particle systems are described by Lagrangians
that, being formulated in the original space-time (so they are intrinsic), in return for they
depend on higher derivatives. Therefore, the attractive point in this new philosophy is that the
spinning degrees of freedom are assumed to be encoded in the geometry of the trajectories. Finally,
the Poincar\'e and invariance requirements imply that the admissible Lagrangian densities must
depend on the extrinsic curvatures of the curves in the background gravitational field.

\vspace{4 mm}

\noindent Most of the published papers, in this direction, involve
actions that only depend on the first curvature of trajectories
(the curvature, which plays the role of proper acceleration of the
particle). However, it seems important to investigate models of
particles with curvature and torsion.

\vspace{4 mm}

\noindent Along this note, $M(C)$, will denote a three dimensional
space with constant curvature $C$. In a suitable space of curves,
$\Lambda$ in $M(C)$, (for example the space of closed curves or
that of curves satisfying certain second order boundary data,
clamped curves), we have a three-parameter family of actions,
$\{\mathcal{F}_{mnp}:\Lambda\rightarrow\mathbb{R} \, : \,
m,n,p\in\mathbb{R}\}$, defined by

\vspace{4 mm}

\begin{equation}
{\cal F}_{mnp}(\gamma)=\int_{\gamma}(m+n\kappa+p\tau)ds,
\end{equation}

\vspace{4 mm}

\noindent where $s$, $\kappa$ and $\tau$ stand for the arclength parameter, curvature and torsion
of $\gamma$, respectively.

\vspace{4 mm}

\noindent The main purpose of this note is to determine, explicitly and completely, the moduli
space of trajectories in the particle model $[M(C),\mathcal{F}_{mnp}]$. In particular, we provide
algorithms to obtain the trajectories of a given model. The closed trajectories, when there exist,
are also obtained from an interesting quantization principle.

\vspace{4 mm}

\noindent It should be noticed that, this problem was considered in \cite{Kuz-Ply} when $C=0$, flat
space. In that paper, the authors showed that trajectories are helices (that is curves with both
curvature and torsion being constant) in $M(0)$. However, this is not true. In fact, we prove here
that trajectories in the model $[M(0),\mathcal{F}_{0np}]$ are curves of Lancret with slope
determined from the values of $n$ and $p$. To understand this note better, we recall in the next
section the nice geometry of the Lancret curves not only in classical setting, $C=0$, but also when
$C$ is arbitrary.

\vspace{4 mm}

\section{The extended Lancret program}

\noindent A curve of Lancret (or general helix) in $\mathbb{R}^3$
is a curve with constant slope, that is, one whose tangent makes a
constant angle with a fixed straight line (the axis of the general
helix). In other words, the tangent indicatrix of a curve of
Lancret lies in a plane of $\mathbb{R}^3$. The two main statements
in the theory of these curves are

\begin{enumerate}
\item A classical result stated by M.A.Lancret in 1802 and first
proved by B. de Saint Venant in 1845 (see \cite{Struik} for
details) which gives an algebraic characterization for Lancret's
curves. {\em The curves of Lancret are those curves that the ratio
of curvature and torsion is constant}. \item The geometric
approach to the problem of {\em solving natural equations} for
general helices in $\mathbb{R}^3$. {\em A curve in $\mathbb{R}^3$
is a Lancret one if and only if it is a geodesic of a right
cylinder on a plane curve}.
\end{enumerate}

\noindent Notice that this class of curves includes not only those curves with torsion vanishing
identically but also the ordinary helices (helices) which have both torsion and curvature being
nonzero constants. These cases correspond, in the geometric approach, with geodesics of right
cylinders shaped on plane curves with constant curvature. A plane with directrix being a straight
line and a circular right cylinder determined by a circle, respectively. We will refer these two
cases as trivial Lancret curves.

\vspace{4 mm}

\noindent In \cite{B0}, the second author used the concept of {\em Killing vector field along a
curve} to define the notion of general helix in a three dimensional real space form, $M(C)$. Then,
he obtained the extension of the Lancret program to this framework. It is a bit more subtle than
one might suppose a priori, as evidenced by the difference between the spherical and the hyperbolic
cases. In fact, while the former case is nicely analogous to the Euclidean one, the later only
presents trivial Lancret curves. To be more precise

\begin{enumerate}
\item A curve in the hyperbolic space, $\mathbb{H}^3$, is a
general helix if and only if either its torsion vanises
identically or it is an ordinary helix. That is, the class of
Lancret curves in a hyperbolic space is just reduced to that of
ordinary Lancret curves. \item A curve in the sphere,
$\mathbb{S}^3$ with constant curvature $C$, is a general helix if
and only if either its torsion vanishes identically or the
curvature, $\kappa$, and the torsion, $\tau$, are related by
$\tau=a\kappa\pm\sqrt{C}$, where $a$ is a certain constant which
will be interpreted as a kind of slope.
\end{enumerate}

\noindent In that paper, the solving natural equations is obtained
as follows. {\em A curve in $\mathbb{S}^3$ is a general helix if
and only if it is a geodesic of a Hopf cylinder}. That is a
surface obtained when one makes the complete lifting, via the
usual Hopf map, of a curve in the corresponding round two sphere.

However, the {\em closed curve problem} for general helices in
$\mathbb{S}^3$ was also given by taking advantage from the well
known isometry type of the Hopf tori obtained on closed
directrices in the two sphere and it gives a very deep difference
with respect to the classical setting.

\vspace{8 mm}

\noindent The main result of this note can be stated as follows

\vspace{4 mm}

\noindent {\bf A curve $\gamma\in\Lambda$ is a critical point of ${\cal F}_{mnp}$ if and only if
$\gamma$ is a Lancret curve in $M(C)$. In other words, the spinning relativistic particles in the
model $[M(C),{\cal F}_{mnp}]$ evolve along Lancret curves of $M(C)$}.

\section{The field equations}

\noindent The metric of $M(C)$ will be denoted by $g=<,>$ and its
Levi-Civita connection by $\nabla$. Let
$\gamma=\gamma(t):I\subset\mathbb{R}\rightarrow M(C)$ be an
immersed curve with speed $v(t)=\mid\gamma^{\prime}(t)\mid$,
curvature $\kappa$, torsion $\tau$ and Frenet frame $\{T,N,B\}$.
Then, one can write the Frenet equations of $\gamma$ as

\begin{eqnarray*}
\nabla_T T & = & \kappa N, \\
\nabla_T N & = & -\kappa T -\tau B, \\
\nabla_T B & = & \tau N.
\end{eqnarray*}

\vspace{4 mm}

\noindent In order to derive first variation formulas for ${\cal
F}_{mnp}$, we will use the following standard terminology (see
\cite{Langer-Singer1} for details). For a curve
$\gamma:[0,L]\rightarrow M$, we take a variation,
$\Gamma=\Gamma(t,r):[0,L]\times
(-\varepsilon,\varepsilon)\rightarrow M$ with
$\Gamma(t,0)=\gamma(t)$. Associated with this variation is the
variation vector field $W=W(t)=\frac{\partial\Gamma}{\partial
r}(t,0)$ along the curve $\gamma(t)$. We also write
$V=V(t,r)=\frac{\partial\Gamma}{\partial t}(t,r)$, $W=W(t,r)$,
$v=v(t,r)$, $T=T(t,r)$, $N=N(t,r)$, $B=B(t,r)$, etc., with the
obvious meanings. We let $s$ denote arclength, and put $V(s,r)$,
$W(s,r)$ etc., for the corresponding reparametrizations. To obtain
the formulas without doing tedious computations, we quote general
formulas for the variations of $v$, $\kappa$ and $\tau$ in
$\gamma$ and in the direction of $W$. These are obtained using
standard computations that involve the Frenet equations

\begin{eqnarray*}
W(v) & = & <\nabla_T W,T>v, \\
W(\kappa) & = & <\nabla^2_T W,N>-2\kappa<\nabla_T W,T>+C<W,N>, \\
W(\tau) & = & \left(\frac{1}{\kappa}<\nabla^2_T W+CW,B>\right)_s+\tau<\nabla_T W,T>+\kappa<\nabla_T W,B>,
\end{eqnarray*}
where the subscript $s$ denotes differentiation with respect to the arclength.

\vspace{4 mm}

\noindent Now, we use a standard argument which involves the above obtained formulas and some
integrations by parts to get the variation of ${\cal F}_{mnp}$ along $\gamma$ in the direction of
$W$

\begin{equation} \delta{\cal F}_{mnp}({\gamma})[W]=\int_{\gamma}<\Omega(\gamma),W>ds+\left[{\cal B}(\gamma,W)\right]^L_0,
\end{equation}
where $\Omega(\gamma)$ and ${\cal B}(\gamma,W)$ stand for the Euler-Lagrange and Boundary
operators, respectively, and they are given by

\[ \Omega(\gamma)=(-m\kappa+p\kappa\tau-n\tau^2+nC)N+(p\kappa_s-n\tau_s)B, \]

\begin{eqnarray*}  {\cal B}(\gamma,W)& = & -\frac{p}{\kappa}<\nabla^2_T W,B>+<\nabla_T W,N> \\
& + & m<W,T>+\left(n\tau-\frac{pC}{\kappa}-p\kappa\right)<W,B>.
\end{eqnarray*}

\vspace{8 mm}

\noindent {\bf Proposition 1 (Second order boundary conditions)} {\em Given $q_1,q_2\in M$ and
$\{x_1,y_1\}$, $\{x_1,y_1\}$ orthonormal vectors in $T_{q_1}M$ and $T_{q_2}M$ respectively, define
the space of curves

\[ \Lambda=\{\gamma:[t_1,t_2]\rightarrow M \, : \, \gamma(t_i)=q_i, T(t_i)=x_i, N(t_i)=y_i, 1\leq i\leq 2\}. \]

\noindent Then, the critical points of the variational problem ${\cal F}_{mnp}:\Lambda\rightarrow
R$ are characterized by the following Euler-Lagrange equations}

\begin{eqnarray}
-m\kappa+p\kappa\tau-n\tau^2+nC & = & 0, \\
p\kappa_s-n\tau_s & = & 0.
\end{eqnarray}

\vspace{4 mm}

\noindent {\bf Proof.} Let $\gamma\in\Lambda$ and $W\in T_{\gamma}\Lambda$, then $W$ defines a
curve in $\Lambda$ associated with a variation $\Gamma=\Gamma(t,r):[0,L]\times
(-\varepsilon,\varepsilon)\rightarrow M$ of $\gamma$,$\Gamma(t,0)=\gamma(t)$. Therefore, we can
make the following computations along $\Gamma$

\begin{eqnarray*}
W & = & d\Gamma(\partial_r), \\
\nabla_T W & = & fT+d\Gamma(\partial_r T), \\
\nabla_T^2 W & = & (\partial_s f+f)T+(\kappa f+\partial_r \kappa)N+\kappa d\Gamma(\partial_r N)+R(T,W)T,
\end{eqnarray*}
here $f=\partial_r(\log{v})$. Then, we evaluate these formulas along the curve $\gamma$ by making
$r=0$ and use the second order boundary conditions to obtain the following values at the endpoints
\begin{eqnarray*}
W(t_i) & = & 0, \\
\nabla_T W(t_i) & = & f(t_i)x_i, \\
\nabla_T^2(t_i) W & = & (\partial_s(f)+f)(t_i)x_i+(\kappa f+\partial_r \kappa)(t_i)y_i.
\end{eqnarray*}

\noindent
As a consequence,
\[ \left[{\cal B}(\gamma,W)\right]^{t_2}_{t_1}=0. \]

\noindent Then, $\gamma$ is a critical point of the variational problem ${\cal
F}_{mnp}:\Lambda\rightarrow R$, that is $\delta{\cal F}_{mnp}({\gamma})[W]=0$, for any $W\in
T_{\gamma}\Lambda$ if and only if $\Omega(\gamma)=0$ which gives (3) and (4).

\vspace{8 mm}

\section{The moduli spaces of trajectories}

\noindent The field equations, (3,4), can be nicely integrated. The set of solutions is summarized
by the following three tables which correspond with Euclidean, hyperbolic and spherical case,
respectively. All the solutions are Lancret curves. Similarly to the Euclidean case, curves with
zero torsion, including geodesics, and helices are considered as special cases of Lancret
curves(trivial Lancret curves). For simplicity of interpretation, we have represented different
cases according with the values of the three parameters that define the action.

\vspace{12 mm}

\begin{tabular}{|l|l|l|p{8cm}|}    \hline
\bfseries $m$ &\bfseries $n$ &\bfseries $p$ &\bfseries Solutions
in $\mathbb{R}^3$, $C=0$\\   \hline
$\neq 0$ & $=0$ & $=0$ &  Geodesics $\kappa=0$\\
$=0$ & $=0$ & $\neq 0$ &  Circles $\kappa$ constant and $\tau=0$\\
$=0$ & $\neq 0$ & $=0$ &  Plane curves $\tau=0$\\
$\neq 0$ & $\neq 0$ & $=0$ & Helices with $\kappa=\frac{-n\tau^2}{m}$\\
$\neq 0$ & $=0$ & $\neq$ 0 & Helices with arbitrary $\kappa$ and $\tau=\frac{m}{p}$\\
$=0$ & $\neq 0$ & $\neq$ 0 & Lancret curves with $\tau=\frac{p}{n}\kappa$\\
$\neq 0$ & $\neq 0$ & $\neq 0$ & Helices with
$\kappa=\frac{-na^2}{m+ap}$, $\tau=\frac{ma}{m+ap}$ and $a\in
\mathbb{R}-\{-\frac{m}{p}\}$ \\\hline
\end{tabular}

\vspace{12 mm}

\begin{tabular}{|l|l|l|p{8cm}|}    \hline
\bfseries $m$ &\bfseries $n$ &\bfseries $p$ &\bfseries Solutions
in $\mathbb{H}^3$, $C=-c^2$\\   \hline
$\neq 0$ & $=0$ & $=0$ &  Geodesics $\kappa=0$\\
$=0$ & $=0$ & $\neq 0$ &  Curves with $\kappa$ constant and $\tau=0$\\
$=0$ & $\neq 0$ & $=0$ &  Do not exist\\
$\neq 0$ & $\neq 0$ & $=0$ & Helices with $\kappa=\frac{-n(c^2+\tau^2)}{m}$\\
$\neq 0$ & $=0$ & $\neq$ 0 & Helices with arbitrary $\kappa$ and $\tau=\frac{m}{p}$\\
$=0$ & $\neq 0$ & $\neq$ 0 & Helices with $\kappa=\frac{-n(c^2+a^2)}{ap}$ and $\tau=-\frac{c^2}{a}$
and $a\in\mathbb{R}-\{0\}$\\
$\neq 0$ & $\neq 0$ & $\neq 0$ & Helices with
$\kappa=\frac{-n(c^2+a^2)}{m+ap}$, $\tau=\frac{ma-pc^2}{m+ap}$ and
$a\in \mathbb{R}-\{-\frac{m}{p}\}$ \\\hline
\end{tabular}

\vspace{12 mm}

\begin{tabular}{|l|l|l|p{8cm}|}    \hline
\bfseries $m$ &\bfseries $n$ &\bfseries $p$ &\bfseries Solutions
in $\mathbb{S}^3$, $C=c^2$\\   \hline
$\neq 0$ & $=0$ & $=0$ &  Geodesics $\kappa=0$\\
$=0$ & $=0$ & $\neq 0$ &  Circles $\kappa$ constant and $\tau=0$\\
$=0$ & $\neq 0$ & $=0$ &  Horizontal lifts, via the Hopf map, of curves in $\mathbb{S}^2$\\
$\neq 0$ & $\neq 0$ & $=0$ & Helices with $\kappa=\frac{n(c^2-\tau^2)}{m}$\\
$\neq 0$ & $=0$ & $\neq$ 0 & Helices with arbitrary $\kappa$ and $\tau=\frac{m}{p}$\\
$=0$ & $\neq 0$ & $\neq$ 0 & Helices with $\kappa=\frac{n(c^2-a^2)}{ap}$ and $\tau=\frac{c^2}{a}$
and $a\in \mathbb{R}-\{0\}$\\
$\neq 0$ & $\neq 0$ & $\neq 0$ & Helices with
$\kappa=\frac{n(c^2-a^2)}{m+ap}$, $\tau=\frac{ma+pc^2}{m+ap}$
and $a\in \mathbb{R}-\{-\frac{m}{p}\}$\\
$\neq 0$ & $\neq 0$ & $\neq 0$ & Lancret curves with
$\tau=\frac{p}{n}\kappa-\frac{m}{p}$ and $c=\pm\frac{m}{p}$
\\\hline
\end{tabular}

\vspace{12 mm}

\noindent Let us make a few remarks on the solutions we have obtained

\begin{enumerate}
    \item {\bf Euclidean case.} The model $[M(0),\mathcal{F}_{m0p}]$ is also related with the total
    twist of a Frenet ribbon, \cite{Tyson+}. The search of trajectories in the models $[M(0),\mathcal{F}_{mn0}]$
    and $[M(0),\mathcal{F}_{m0p}]$, under the additional assumpion that they are constrained to lie on
    a given surface, had been previously considered by L.Santal\'o, \cite{Santalo}. However, only
    in the case where such a surface is a round sphere the solution is clear.

    From the table corresponding to $\mathbb{R}^3$, we see that the model of higher interest is
    $[M(0),\mathcal{F}_{0np}]$. An algorithm to obtain explicitly, up to motions in $\mathbb{R}^3$, all the
    trajectories of this model works as follows
    \begin{enumerate}
        \item We take a plane, say $\Pi$ in $\mathbb{R}^3$ and a curve, say $\gamma(u)$, $u\in I\subset\mathbb{R}$, contained in $\Pi$.
        \item Let $\xi$ be a unit vector orthogonal to $\Pi$ and denote by $\mathbf{C}_{\gamma}$
        the right cylinder shaped from $\gamma(u)$, that is, the image in $\mathbb{R}^3$ of the map
        $\phi:I\times\mathbb{R}\rightarrow\mathbb{R}^3$ defined by

        \[ \phi(u,v)=\gamma(u)+v \, \xi. \]
        \item Let $\gamma_{np}(t)$ be a geodesic of $\mathbf{C}_{\gamma}$ with slope $\theta$,
        $\tan{\theta}=\frac{n}{p}$, that is

        \[\gamma_{np}(t)=\phi(nt,pt)=\gamma(nt)+pt \, \xi\]
        then, $\gamma_{np}(t)$ is a  trajectory of the model $[M(0),\mathcal{F}_{0np}]$.
        \item Moreover, all the trajectories of particles in this model can be obtained in this way.
        Consequently, up to motions in $\mathbb{R}^3$, the set of trajectories or the moduli space of
        curves that are solutions to the field equations in the model $[M(0),\mathcal{F}_{0np}]$
        can be identified with the space

        \[ \Gamma_{np}=\{\gamma_{np} \, : \, \gamma \, {\rm is \, a \, curve \, in} \, \Pi\}.\]
    \end{enumerate}

    \vspace{4 mm}

    \item {\bf Hyperbolic case.} This is the uninteresting case because the hyperbolic space is
    free of non-trivial Lancret curves. Therefore, most of the models $[M(-c^2),\mathcal{F}_{mnp}]$
    admit a one-parameter family of trajectories which are trivial Lancret curves or helices. The
    exception to this rule is the model $[M(-c^2),\mathcal{F}_{0n0}]$. That is, this associated
    with the action measuring the total curvature of trajectories which does not provide any
    consistent dynamics, (see \cite{ABG} for more details).
    \item {\bf Spherical case.} The most interesting models in the sphere are either
    $[M(c^2),\mathcal{F}_{0n0}]$ and $[M(c^2),\mathcal{F}_{mnp}]$ with $m.n.p\neq 0$. The former
    one corresponds again with the action giving the total curvature. In \cite{ABG}, it is showed
    that the three-dimensional sphere is the only space (no matter the dimension) with constant
    curvature providing a consistent dynamics for this action. More precisely, the trajectories of
    this model are nothing but the horizontal lifts, via the usual Hopf map, of arbitrary curves in
    the two-sphere. It should be noticed that those curves are Lancret ones where the curvature is an
    arbitrary function while the torsion is nicely determined by the radius of the three-sphere.
    The later case provides a model which has two kinds of trajectories. First, it has a
    one-parameter class of trajectories, $\mathcal{T}$, which are helices and no more comments on it (see table 3).
    However, the dynamics of this model is completed with a second class of trajectories, $\mathcal{T}_{mnp}$,
    that are Lancret curves with
    \[\tau=\frac{p}{n} \, \kappa-\frac{m}{p}.\]

    \noindent For a better understanding of the family of trajectories $\mathcal{T}_{mnp}$, we will design an
    algorithm to obtain its geometric integration.

    \begin{enumerate}
        \item First of all, notice that the ratio $\frac{m}{p}$ and the radius, $r$, of
        the three-dimensional sphere are constricted to satisfy
        \[\frac{p}{m}=\pm r,\]
        therefore, without loss of generality we may assume that $r=1$ and so $m=\pm p$, we will put $m=p$
        in the discussion.
        \item Let consider the usual Hopf map, $\pi:\mathbb{S}^3(1)\rightarrow\mathbb{S}^2(\frac{1}{2})$, between round spheres
        of radii $1$ and $\frac{1}{2}$, respectively. In this setting, $\pi$ is a Riemannian submersion and the
        flow of geodesic fibres is generated by a Killing vector field, $\eta$, which is sometimes called the
        {\em Hopf vector field}.
        \item  If $\beta(u)$, $u\in I\subset\mathbb{R}$, is a curve in $\mathbb{S}^2(\frac{1}{2})$,
        then $\mathbf{H}_{\beta}=\pi^{-1}(\beta)$ is a flat surface of $\mathbb{S}^3(1)$ called the {\em Hopf tube}
        on $\beta$. If $\bar{\beta}$ is a horizontal lift of $\beta$, one can use the natural action of the unit
        circle on $\mathbb{S}^3(1)$ to see that the map, $\psi:I\times\mathbb{R}\rightarrow \mathcal{H}_{\beta}$, defined by

        \[\psi(u,v)=e^{iv}\bar{\beta}(u),\]
        is a Riemannian covering map which carries coordinate curves in horizontal lifts of $\beta$ and
        fibres, respectively.
        \item Let $\beta_{np}(t)$ be a geodesic of $\mathbf{H}_{\beta}$ with slope $\theta$,
        $\tan{\theta}=\frac{n}{p}$, that is
        \[\beta_{np}(t)=\psi(nt,pt)=e^{ipt}\bar{\beta}(nt)\]
        then, $\beta_{np}(t)$ is a Lancret curve in $\mathbb{S}^3(1)$, which is a trajectory in $\mathcal{T}_{mnp}$.
        \item The converse also holds. Every trajectory in $\mathcal{T}_{mnp}$ can be regarded as a
        geodesic, with slope $\frac{n}{p}$, in a Hopf tube, $\mathbf{H}_{\beta}=\pi^{-1}(\beta)$,
        shaped on a curve, $\beta$ in $\mathbb{S}^2(\frac{1}{2})$. Consequently,

        \[ \mathcal{T}_{mnp}=\{\beta_{np} \, : \, \beta \, {\rm is \, a \, curve \, in} \, \mathbb{S}^2(\frac{1}{2})\},\]
        recall that $m=\pm p$.
        \item Since the slope, $\frac{n}{p}$, is known once we choose the action, the space of
        trajectories, $\mathcal{T}_{mnp}$, is completely determined, up to congruence, when we give
        the curvature, in $\mathbb{S}^2(\frac{1}{2})$, of curves $\beta$.
        \item The conclusion is that the space of trajectories and so the dynamics of the
        particle system $[M(c^2),\mathcal{F}_{mnp}]$ with $m.n.p\neq 0$, is

        \[\mathcal{T}\cup\mathcal{T}_{mnp},\]
        and so the moduli space of solitons is defined by a couple of parameters, a real number
        fixing the helix in $\mathcal{T}$ and a smooth function in $C^\infty(I,\mathbb{R})$ which works as
        the curvature function of a curve in $\mathbb{S}^2(\frac{1}{2})$ which determines the Hopf tube and
        so the corresponding geodesic with slope $\frac{n}{p}$.
    \end{enumerate}

    \end{enumerate}

\section{Closed trajectories}

To study closed trajectories, we will modify a little bit the model $[M(C),\mathcal{F}_{mnp}]$ in
the sense that the action, $\mathcal{F}_{mnp}$, is now assumed to be defined on the space of closed
curves in $M(C)$. In this case no boundary conditions are necessary. For obvious reasons, we will
restrict ourselves to the spherical case and without loss of generality we will consider the sphere
of radius one. Then, we have similar field equations

\vspace{4 mm}

\noindent {\bf Proposition 2.} {\em Let ${\cal C}$ be the space of
immersed closed curves in $\mathbb{S}^3(1)$. The critical points
of the variational problem associated with the action ${\cal
F}_{mnp}:{\cal C}\rightarrow\mathbb{R}$ are those closed curves
that are solutions of the following Euler-Lagrange equations

\begin{eqnarray}
(p\tau-m)\kappa+n(1-\tau^2) & = & 0,\\
p\kappa_s-n\tau_s & = & 0.
\end{eqnarray}}

\vspace{4 mm}

\noindent It is obvious that the solutions of the above stated
field equations are Lancret curves in $\mathbb{S}^3(1)$.
Consequently, we need to determine closed Lancret curves in
$\mathbb{S}^3(1)$. These trajectories can be characterized
according to the following algorithm
\begin{enumerate}
    \item If we choose a closed curve, $\beta(u)$, $u\in\mathbb{R}$, in $\mathbb{S}^2(\frac{1}{2})$,
        then, its Hopf tube, $\mathbf{H}_{\beta}=\pi^{-1}(\beta)$, turn to a flat torus of $\mathbb{S}^3(1)$.
    \item The isometry type of a Hopf torus can be determined using the Riemannian covering map,
    $\psi:\mathbb{R}^2\rightarrow \mathbf{H}_{\beta}$, and some well known machinery (see \cite{Greub}, Vol II,
    p.293, for details and also \cite{Pinkall}). In fact, $\mathbf{H}_{\beta}=\pi^{-1}(\beta)$ is isometric
    to the $\mathbb{R}^2/R$, where $R$ is the lattice in $\mathbb{R}^2$ generated by $(2A,L)$ and $(2\pi,0)$. Here $L$
    denotes the length of $\beta$ and $A\in (-\pi,\pi)$ the oriented area enclosed by $\beta$ in the two sphere.
    \item Consequently, a Lancret curve of $\mathbb{S}^3(1)$ (recall a geodesic of
    $\mathbf{H}_{\beta}=\pi^{-1}(\beta)$) closes if and only if its inverse slope, $\omega=\cot{\theta}$, satisfies
    \[ \omega= \frac{1}{L}(2A+q\pi), \]
    where $q$ is a rational number.
    \newpage
    \item On the other hand, $\gamma\in {\cal C}$ is a trajectory of $[M(C),\mathcal{F}_{mnp}]$ if and only if its
    inverse slope satisfies $\omega=\frac{p}{n}$. In particular, it closes if and only if its inverse slope,
    $\omega=\frac{p}{n}$, satisfies the following quantization principle

    \[ \frac{p}{n} \, L-2A \qquad {\rm is \quad a \quad rational \quad multiple \quad of} \quad \pi.\]
    \end{enumerate}

\section{Existence of closed trajectories}

\noindent For simplicity, we can assume the area $A$ to be
positive, changing if necessary the orientation of $\beta$. The
only further restriction on $(A,L)$ to define an embedded closed
curve in the two sphere is given by the iso-perimetric inequality
in $\mathbb{S}^2(\frac{1}{2})$, which can be written as

\[ L^2+4A^2-4\pi A\geq 0. \]

\noindent In terms of $(2A,L)$, this inequality is expressed as

\[ L^2+(2A-\pi)^2\geq\pi^2.\]

\vspace{4 mm}

\noindent In the $(L,2A)$-plane, we define the region
\[ \Delta=\{(L,2A) \, : \, L^2+(2A-\pi)^2\geq\pi^2 \, {\rm and} \, 0\leq A\leq\pi\}, \]

\noindent then for each point $z\in(L,2A)\in\Delta$ there is an
embedded closed curve, $\beta^{z}$, in $\mathbb{S}^2(\frac{1}{2})$
with length $L$ and enclosed area $A$. We already know that a
geodesic, $\beta_{np}^{z}$, of
$\mathbf{H}_{\beta^{z}}=\pi^{-1}(\beta^{z})$ with slope
$\frac{1}{\omega}=\frac{n}{p}$ is a trajectory of the model
$[M(C),\mathcal{F}_{mnp}]$. Moreover, we use the quantization
principle to see that it closes if and only if the straight line,
in the $(L,2A)$-plane, with slope $\omega=\frac{p}{n}$ cuts the
$2A$-axis at a height which is a rational multiple of $\pi$.

\vspace{4 mm}

\noindent {\bf Theorem.} {\em For any couple of parameters, $n$ and $p$ with $n.p\neq 0$, there
exists an infinite series of closed trajectories in the model $[M(C),\mathcal{F}_{mnp}]$ on the
three-dimensional sphere of radius $r=\frac{p}{m}$, $m\neq 0$. This series includes all the
geodesics $\beta_{np}^z$ in $M_{\beta^z}=\pi^{-1}(\beta^z)$ with slope $\omega=\frac{p}{n}$ and
$\beta^z$ determined as above by $z=(L,2A)$ in the following region}

\[ \Delta\bigcap\left( \bigcup \, (\frac{p}{n} \,  L-2A=q\pi)\right)_{{q\in\mathbb{Q}}}. \]

\newpage

\noindent {\bf Remark.} A quantization principle to characterize
the moduli sub-space of closed trajectories in $\mathcal{T}$ can
be also obtained. In this case, since the trajectories are
helices, then they are geodesics of Hopf tubes (Hopf tori to be
closed) shaped on geodesic circles in $\mathbb{S}^2(\frac{1}{2})$.
Moreover the slope in the corresponding flat torus depends on the
parameter $a$ (see Table 3) according to

\[ \omega=\frac{p-m}{n(1+a)}.\]

\section{Conclusions}

\noindent The events of this note take place in $D=3$ spaces with constant curvature, $M(C)$. In
this setting, we have considered models for relativistic particles where the Lagrangian densities
depend linearly on both the curvature and the torsion of the trajectories.

\noindent The moduli spaces of classical solutions are completely and explicitly obtained. A part
of these spaces in flat backgrounds was known, solutions being helices. However, the more
interesting models are those where non helicoidal solutions appear. In these cases the solutions
are non trivial Lancret curves in flat spaces and spherical ones, respectively.

\noindent  The complete spaces of solutions are formally described in three tables. However, we
design algorithms providing the geometric integrations of these spaces of solutions. The geometry
of Lancret curves in the classical setting so as its extension to spherical framework, based in the
the Hopf map, are the chief points in these algorithms.

\vspace{8 mm}

\noindent {\bf Acknowledgments}

\noindent This research has been partially supported by a MCYT and FEDER grant no BFM2001-2871-C04.

\vspace{8 mm}

\noindent
J.Arroyo and O.J.Garay: Departamento de Matem\'aticas\\
Universidad del Pais Vasco/Euskal Herriko Unibertsitatea\\
Aptdo. 644, 48080 Bilbao, Spain\\
email: mtparolj@lg.ehu.es (J.Arroyo), mtpgabeo@lg.ehu.es
(O.J.Garay)

\vspace{4 mm}

\noindent
M.Barros: Departamento de Geometr\'\i{}a y Topolog\'\i{}a.Universidad de Granada\\18071, Granada. Spain\\email:mbarros@goliat.ugr.es

\end{document}